\documentclass[12pt,epsf,graphicx,epsfig]{elsarticle}
\usepackage{epsf}
\usepackage{epsfig}
\usepackage{graphicx}
\usepackage{amsmath}
\usepackage{amsfonts,amsbsy}
\usepackage{amssymb}
\usepackage{graphicx}
\usepackage{bm}
\usepackage{resizegather}
\usepackage[colorlinks=true,pdfstartview=FitV,linkcolor=red,citecolor=blue,urlcolor=blue]{hyperref}

\def\be{\begin{equation}}
\def\ee{\end{equation}}
\def\bea{\begin{eqnarray}}
\def\eea{\end{eqnarray}}

\def\vp#1{\vec{#1}_\perp}
\def\d{\dagger}
\def\lrd{\overset{\leftrightarrow}{\partial_0}}
\def\lrdtau{\overset{\leftrightarrow}{\partial_\tau}}

\newcommand{\pvec}[1]{\vec{#1}\mkern2mu\vphantom{#1}}
\def\pvp#1{\pvec{#1}'_\perp}

\mathchardef\Re="023C
\mathchardef\Im="023D

\begin{document}

\title{\bf Odd Azimuthal Anisotropy of the Glasma for $pA$ Scattering}

\author[int,ccnu]{Larry McLerran}
\author[rbrc]{Vladimir Skokov}

\address[int]{Institute for Nuclear Theory, University of Washington, Box 351550, Seattle, WA, 98195, USA}
\address[ccnu]{China Central Normal University, Wuhan, China}
\address[rbrc]{RIKEN/BNL, Brookhaven National Laboratory,  Upton, NY 11973}

\begin{abstract}
In this paper we analytically extract the odd azimuthal anisotropy in the
Classical Yang--Mills equations for the Glasma for pA collisions.  We compute
the first non-trivial
term in the expansion of the proton sources of color charge.  The computation
is valid in the limit of a large nucleus when the produced particle momenta are
larger than the saturation momentum of the proton.
\end{abstract}

\maketitle

\section{Introduction}
Computations using the theory of the Color Glass Condensate can generate even
flow harmonics  from initial state
correlations~\cite{Dumitru:2008wn,Dumitru:2010iy,Dusling:2012iga,Dusling:2012wy}.
These correlations are non-vanishing in the limit of an infinite  number of color
sources, but suppressed by the number of colors.  This is in
distinction from
fluctuations generated by a finite number of  scattering centers which are
non-vanishing in the limit of a large number of colors but vanish in the limit of an
infinite number of
color sources~\cite{Alver:2006wh,Alver:2008zza,Bzdak:2013rya,Yan:2014afa,Bzdak:2013raa,Dumitru:2014yza,McLerran:2015sva,McLerran:2016ivs}.
Four- and higher order-particle elliptical anisotropies also demonstrate a
non-trivial behavior as a function of number of colors and number of
sources~\cite{Dumitru:2014yza,Skokov:2014tka,Dumitru:2015cfa}.

The situation for odd harmonics is very interesting.
Unlike the case for even harmonics, to obtain
odd harmonics at small $x$ requires final state interactions
at least in the classical approximation to the
CGC.\footnote{
As demonstrated in Ref.~\cite{Kovner:2016jfp},
odd azimuthal anisotropy is present
in the CGC wave function beyond the classical approximation.}
This is a
consequence of time reversal invariance.  In models of the Glasma, where classical equations are computed
numerically, one
sees odd harmonics, and they indeed develop after the collision
of the nuclei has taken place while final state interactions are in
play~\cite{Gale:2012rq,Lappi:2015vta,Schenke:2015aqa}.

It is the purpose of this paper to elaborate somewhat on the generation of flow in the classical
equations that
describe the evolution of the Glasma~\cite{Kovner:1995ja,Kovner:1995ts}
and to build a bridge between the analytical calculations in the dilute--dense
limit (see e.g.~Ref.~\cite{Kovner:2010xk,Kovner:2012jm})  and the dense--dense numerical
results~\cite{Lappi:2009xa,Schenke:2015aqa}.
We begin by solving the classical Yang--Mills equations around
the free field equations for a distribution
of fluctuating sources.
We consider a proton nucleus collision in a momentum range where the
field of the proton can be treated as weak.
We show explicitly that there are no odd harmonics generated by this lowest order
solution.  We then iterate the equations
around the leading order, treating the color field of the
nucleus to all orders, and we find that we generate odd moments of azimuthal anisotropy in the first
such iteration of these equations.   The  non-zero contribution to odd
harmonics arises from the interference  of the leading  and next-to-leading
orders.  We find remarkable
simplification
for the result for such odd moments when gluons are put on mass shell, and we integrate over
intermediate coordinates associated with iterating the equations.

This exercise is not only academic, in that it provides analytic confirmation of what is already
known from numerical simulation, but it is  also useful for describing dilute
systems such
provided by pp and pA collisions, since the analytical form may be somewhat
simpler to use than numerical solutions to the full scattering problem.

\section{Notation and review of known results}
\label{Sec:Notation}

In this section we set up our notation, and will use well known results from the literature concerning
the classical equations that describe the Color Glass Condensate and the Glasma \cite{Gelis:2010nm}.
We begin by writing down the color field of an isolated nucleon or nucleus as
\begin{equation}
	\alpha^i_{m}(x_\perp) = - \frac{1}{ig} U_m(\vp{x}) \partial^i U^\d_m(\vp{x}) =  \frac{1}{ig} [\partial^i U_m(\vp{x}) ] U^\d_m(\vp{x}),
\end{equation}
where the Wilson lines  are in the fundamental representation.
The field is generated by valence color charges
\begin{equation}
	\partial_i \alpha_i(\vp{x}) = g \rho(\vp{x})\;.
	\label{Eq:WWfield}
\end{equation}
The label $m$ is $1$ for the field of a proton and $2$ for the nucleus.
This is the field that describes the nucleon or nucleus before the
collision, and is the same as for an isolated nucleon or nucleus.
We consider the source for the proton  $\rho_1$ to be weak and expand the corresponding
Wilson lines into power series to get
\begin{equation}
	\alpha^i_1(\vp{x})  =  \partial^i \Phi_1 (\vp{x}) - \frac{ig}{2}
	\left(
	\delta_{ij} - \frac{\partial_i \partial_j}{\partial^2}
	\right) \left[ \partial^j \Phi_1 (\vp{x}), \Phi_1 (\vp{x})   \right] + {\cal O} (\Phi_1^3)  \;,
	\label{Eq:U1_expansion}
\end{equation}
where
\begin{equation}
	\Phi_1(\vp{x})  = \frac{g}{\partial^2}  \rho(\vp{x})\;.
	\label{Eq:Phi_1}
\end{equation}

The only Wilson lines we will
encounter in the text correspond to the strong, nucleus field $m=2$.
In order to simplify the notation we will omit the redundant subscript, that is
\begin{equation}
	U(\vp{x}) \equiv U_2(\vp{x})\,.
\end{equation}
In what follows we will perform
the expansion in the weak field; the notation for the expansion coefficients is
defined by
\begin{equation}
	f(k) = \lim_{N\to\infty} \sum_{n=1}^N f^{(n)}(k) \,.
\end{equation}
We warn the reader that there might be a potential confusion with the notation of the Hankel functions, which also involves bracketed integers in the superscript.

In this paper we will consider only first two nontrivial corrections,
i.e. we terminate the expansion at $N=2$.  When one computes the cross section
for particle production, one evaluates a square of the amplitudes associated with the color field.
The
first non-trivial correction to particle production that involves a second order iteration of the proton field, is of the order
$\rho_1^3$ and originates from the interference of the leading and next-to-leading order expansion coefficients.
We will argue that this  is the leading order correction  contributing  to the odd
azimuthal anisotropy of double inclusive gluon production.

Following a widely accepted convention,  we define
\begin{equation}
	\tau = \sqrt{2 x_+ x_-},
\end{equation}
where
\begin{equation}
	x_\pm = \frac{x_0\pm x_z}{\sqrt{2}}\,.
\end{equation}

When it is convenient we will use the Milne metric, or, the $\tau-\eta$-coordinates.
In this case the Minkowski coordinates are parametrized by
$$ x = (\tau \cosh \eta, \vec{x}_\perp, \tau \sinh \eta).$$
Here  $\vec{x}_\perp$ is a two-dimensional vector.

We work in the Fock-Schwinger gauge $$A_\tau = x_- A_+ + x_+ A_- = 0.$$
Therefore the $\eta$ component of the vector potential is
\begin{equation}
	A_\eta = x_- A_+  - x_+ A_- = \tau^2 \alpha,
\end{equation}
where $\alpha$ is introduced for convenience.

Since the quantum corrections are explicitly ignored in our classical Yang-Mills
approach,
the field created in collisions is
$\eta$-independent.

For Bessel (Neumann) functions of $n$-th order the following notation is used
$J_n(x)$ ($Y_n(x)$).

\section{Equations of motion}
In the upper light cone, assuming independence of rapidity,
the Classical Yang--Mills (CYM)  equations
$[D_\mu, F^{\mu\nu}] =0$ can be written as
\begin{eqnarray}
	&&\frac{1}{\tau} \partial_\tau \tau \partial_\tau A_i -
	\partial_j (\partial_j \partial A_i - \partial_i \partial A_j)
	\notag \\
	&&\quad + ig \left( \partial_i [A_j, A_i] + \frac{1}{\tau^2} [A_\eta, F_{\eta i}] + [A_j , F_{ij}]
	\right) = 0,  \nonumber \\
	&&\partial_\tau \tau^{-1} \partial_\tau A_\eta
	- \frac{1}{\tau} \partial_\perp^2 A_\eta
	+ ig \left(
	\frac{1}{\tau} \partial_j [A_j, A_\eta] + \frac{1}{\tau} [A_j, F_{j\eta}]
	\right) =0,  \nonumber\\
	&& \partial_\tau \partial_i A_i -ig
	\left(
	\frac{1}{\tau^2} [A_\eta, \partial_\tau A_\eta] +  [A_i, \partial_\tau A_i]
	\right)
	= 0.
\end{eqnarray}
Note that the $\tau$ derivatives  in the second equation can be written
in the following equivalent form
\begin{equation}
	\partial_\tau \tau^{-1} \partial_\tau A_\eta \equiv
	\tau \partial_\tau^2 \alpha + 3 \partial_\tau \alpha \equiv \tau^{-2} \partial_\tau \tau^3 \partial_\tau \alpha.
\end{equation}

\section{Solutions of CYM to leading order in weak field}
In this section we review  the result of Ref.~\cite{Kovner:1995ja,Kovner:1995ts,Dumitru:2001ux} using the notation we defined in Sec.~\ref{Sec:Notation} and present them  in the form
that will be useful for what follows.

The gluon field has the following dependence
\begin{align}
	A^\pm(x^+,x^-,\vec{x}_\perp) &= \pm x^\pm \alpha(\tau, \vec{x}_\perp) \theta(x^+) \theta(x^-),\\
	A^i(x^+,x^-,\vec{x}_\perp)  &=  \alpha^i(\tau, \vec{x}_\perp) \theta(x^+) \theta(x^-) \notag
  \\ &+
	\alpha^i_1( \vec{x}_\perp) \theta(- x^+) \theta(x^-) +
\alpha^i_2(\ \vec{x}_\perp) \theta(x^+) \theta(- x^-)
\end{align}
with the initial conditions obtained by matching the singularities on the light cone
\begin{eqnarray}
	\alpha (\tau\to 0, \vp{x}) &=& \frac{ig}{2}
	[\alpha_1^{i}(\vp{x}) ,\, \alpha_2^{i}(\vp{x})], \\
	\alpha^i (\tau\to 0, \vp{x}) &=&
	\alpha_1^{i}(\vp{x})
	+
	\alpha_2^{i}(\vp{x}).
\end{eqnarray}

The gauge rotation
\begin{eqnarray}
\alpha(\tau,\vec{x}_\perp)
&=&
U(\vec{x}_\perp) \beta (\tau,\vec{x}_\perp) U^\dagger (\vec{x}_\perp)\;, \\
\alpha_i(\tau,\vec{x}_\perp)
&=&
U(\vec{x}_\perp) \left( \beta_i (\tau,\vec{x}_\perp)  -
\frac{1}{ig} \partial_i  \right) U^\dagger (\vec{x}_\perp)
\end{eqnarray}
enables us to perform a systematic expansion in powers of $\rho_1$.

At the leading order, the CYM equations are
\begin{eqnarray}
	&&\left[  \partial_\tau^2
	+ \frac{3}{\tau} \partial_\tau
- \partial_\perp^2 \right] \beta^{(1)}(\tau, \vp{x}) = 0 , \\
&&\partial_\tau \partial_i \beta_i^{(1)}(\tau, \vp{x}) =0 , \\
&&\left[
	\delta^{i j} \left( \partial_\tau^2 + \frac{1}{\tau} \partial_\tau
	- \partial_\perp^2 \right) + \partial_i \partial_j
\right] \beta^{(1)}_j (\tau, \vp{x}) =0
\end{eqnarray}
with solutions
\begin{eqnarray}
	\beta^{(1)} (\tau, \vp{k} ) &=& b_1 (\vp{k}) \frac{J_1(k_\perp \tau)}{k_\perp \tau}, \\
	\beta^{(1)}_i  (\tau, \vp{k} ) &=&
	i \frac{\varepsilon^{ij} k_j}{k_\perp^2} b_2 (\vp{k}) J_0( k_\perp \tau)
	+ i k_i \Lambda(\vp{k})\,.
\end{eqnarray}
The newly introduced functions are defined by the initial conditions
\begin{eqnarray}
	\label{Eq:b1}
	b_1(\vp{x}) &=& ig  U^\d (\vp{x})
	[ \alpha_1^{(1)\,i}(\vp{x}), \alpha_2^i(\vp{x})]  U(\vp{x}) , \\
	\label{Eq:b2}
	b_2(\vp{x})  &=& \epsilon^{ij} \partial^j \left(  U^\d (\vp{x})
	\alpha_1^{(1)\,i}(\vp{x}) U(\vp{x})  \right), \\
	\Lambda(\vp{x}) &=& \frac{\partial^i}{\partial_\perp^2}
	\left(
	U^\d (\vp{x}) \alpha_1^{(1)\,i}(\vp{x}) U(\vp{x})
	\right).
\end{eqnarray}
Note that these functions are manifestly real (to be precise the components), thus the following holds for their Fourier images
\begin{equation}
	f(\vp{k}) = f^*(-\vp{k})\;,
\end{equation}
where
$f(\vp{k})$ is either of $b_1(\vp{k})$, $b_2(\vp{k})$ or $\Lambda(\vp{k})$.

Equations \eqref{Eq:b1} and \eqref{Eq:b2} can also be rewritten in a similar form.
For this, we use the definition of $\alpha_2^i(\vp{x})$ and simplify the commutator in
Eq.~\eqref{Eq:b1}:
~\eqref{Eq:b1}:
\begin{align}
	\label{Eq:b1_sym}
	b_1(\vp{x}) &= \delta^{ij} \Omega_{ij}, \\
	b_2(\vp{x})  &= \epsilon^{ij}
	\Omega_{ij}\;
	\label{Eq:b2_sym}
\end{align}
with
\begin{align}
\notag	\Omega^{ij} (\vp{x}) &=  \left(\alpha_{1 }^{(1)\,i}(\vp{x})\right)_a  \partial^j \left(
	U^\d (\vp{x})
	t_a U(\vp{x})     \right)  \\ &=
	g \left[ \frac{\partial_i}{\partial^2} \rho^a_1(\vp{x}) \right] \partial^j W_{ba} (\vp{x}) t^b,
	\label{Eq:A}
\end{align}
where we used the adjoint Wilson line
$$W_{ab} (\vp{x}) = 2\ {\rm tr} \left( U^\dagger (\vp{x}) t_b U(\vp{x}) t_a \right) .$$
To derive these equations we have made explicit use of the form of the solution
for $\alpha_1^i$ when expanded to first order in the strength of the proton
source, see Eq.~\eqref{Eq:U1_expansion}.

\section{Particle production}
\subsection{LSZ}
We start this section from reviewing the Lehmann--Symanzik--Zimmermann (LSZ)
reduction formula for a scalar field.
The time-dependent creation operator describing one particle at state $\vec{k}$
is defined by
\begin{equation}
	a^+ (\vec{k}, t) =
	\frac{1}{i} \int d^3x \exp(-i k \cdot x)
	\lrd \phi(x) \;,
\end{equation}
where $k$ and $x$ are four-dimensional vectors and $k \cdot x = k_\mu x^\mu$.

We can construct the combination
\begin{align}
	&a^+(\vec{k}, t\to\infty)
	- a^+(\vec{k}, t \to 0) =
	\frac{1}{i} \int_0^\infty dt \partial_0 \left(
	\int d^3x \exp(-i k \cdot x)
	\lrd \phi(x)
	\right) \notag \\ &=
	\frac{1}{i}
	\int_0^\infty dt d^3x \exp(-i k \cdot x)
	\left( \Box + m^2 \right) \phi(x),
	\label{Eq:LSZaa}
\end{align}
where usually instead of $t\to 0$ the limit $t\to-\infty$ is used for the second
term. We chose the limit $t\to 0$ to mimic our problem where
 the initial conditions are formulated on the light
cone.
From the equality~\eqref{Eq:LSZaa}, we can express the creation operator in the
final state by
\begin{align}
	a^+(\vec{k},\infty)
	&=
	\left[
	\frac{1}{i}
	\int d^3x \exp(-i k \cdot x)
\lrd \phi(x)\right]_{t=0}
	\notag
	\\
	&+
	\frac{1}{i}
	\int_0^\infty dt \int d^3x \exp(-i k \cdot x)
	\left( \Box + m^2 \right) \phi(x) \,.
\end{align}
Therefore under the classical approximation,
we  deduce that number of
produced particles  is given by
\begin{align}
	E_k \frac{d N}{d^3 k} = \frac{1}{2 (2\pi)^3}
	&\left|
	\left[ \int d^3x \exp(-i k \cdot x)
	\lrd \phi(x) \right]_{t=0}
	\right. \notag \\ & \left.
	 +
	\int_0^\infty dt \int d^3x \exp(-i k \cdot x)
	\left( \Box + m^2 \right) \phi(x)
	\right|^2\,.
	\label{Eq:LSZ}
\end{align}
Here we have two distinct contributions. One is from the initial time $t=0$
``surface'' and the other involving the time integration from the ``bulk''.
Anticipating the results, we want to comment that the surface contribution
is manifestly $T$-even and thus is not expected to produce non-zero
odd azimuthal anisotropy.

\subsection{Milne metric}
A straightforward generalization of Eq.~\eqref{Eq:LSZ} for $\beta_i$ and $\beta$
in the Milne metric reads
\begin{equation}
	E_k \frac{d N}{d^3 k}  =\frac{1}{8 \pi} \left[
	 \left|
	\mathfrak{S}_\perp (\vp{k})
	+
	\mathfrak{B}_\perp (\vp{k})
	\right|^2
	+ \left|
	\mathfrak{S}_\eta (\vp{k})
	+
	\mathfrak{B}_\eta (\vp{k})
	\right|^2
	\right]\;,
\end{equation}
where the surface contributions at $\tau \to 0+$ are given  by
\begin{align}
	\mathfrak{S}_\perp (\vp{k})  &= \lim_{\tau \to 0+}
	\left(
	\tau H_0^{(1)} (k_\perp \tau) \lrdtau \beta_{\perp} (\tau, \vp{k})
	\right)\;,\\
	\mathfrak{S}_\eta (\vp{k})  &=
\lim_{\tau \to 0+ }
	\left(
	\tau^3 \left\{  \frac{H_1^{(1)}(k_\perp \tau)}{\tau} \lrdtau \beta (\tau, \vp{k})
	\right\}
	\right)\,.
	\label{Eq:Surface3}
\end{align}

The bulk  contributions from the upper light cone are
\begin{align}
	\label{Eq:Bulk_perp}
	\mathfrak{B}_\perp (\vp{k}) &= \int_0^\infty d \tau \tau
	H_0^{(1)} (k_\perp \tau)
	\left\{
		\frac{1}{\tau} \partial_\tau \tau \partial_\tau \beta_{\perp}(\tau, \vp{k}) -
		\partial_\perp^2 \beta_{\perp}(\tau, \vp{k})
	\right\}\;, \\
	\mathfrak{B}_\eta (\vp{k}) &=
	\int_0^\infty d \tau \tau^2
	H_1^{(1)} (k_\perp \tau)
	\left\{
		\frac{1}{\tau^3} \partial_\tau \tau^3 \partial_\tau \beta(\tau, \vp{k}) -
		\partial_\perp^2 \beta(\tau, \vp{k})
	\right\}\; ,
	\label{Eq:Bulk}
\end{align}
where  the transverse part of the field $\beta_i$ is defined as~\footnote{$\epsilon_{ij}$ stands for the antisymmetric tensor, $\epsilon_{12}=1$.}
\begin{equation}
	\beta_\perp (\tau, \vp{k}) = i \frac{\epsilon^{ij}k_j}{k_\perp} \beta_i (\tau, \vp{k}) .
\end{equation}
The imaginary unit is included to guarantee that the function $\beta_\perp (\tau, \vp{x})$ is real.

In order to simplify the notations we will introduce the following combinations
\begin{eqnarray}
	\label{Eq:j_perp_k}
	j_\perp (\tau, \vp{k}) &=& 	\frac{1}{\tau} \partial_\tau \tau \partial_\tau \beta_\perp(\tau, \vp{k}) -  \partial_\perp^2 \beta_\perp(\tau, \vp{k}), \\
	j_i (\tau, \vp{k}) &=& 	\frac{1}{\tau} \partial_\tau \tau \partial_\tau \beta_i(\tau, \vp{k}) -  \partial_\perp^2 \beta_i(\tau, \vp{k}), \\
	j (\tau, \vp{k}) &=& 	\frac{1}{\tau^3} \partial_\tau \tau^3 \partial_\tau \beta(\tau, \vp{k}) -  \partial_\perp^2 \beta(\tau, \vp{k})\;.
\end{eqnarray}
which will be referred to as
``currents'' because they vanish in the absence of non-trivial
interaction in the bulk.

\subsection{Absence of odd azimuthal anisotropy at leading order}
Lets consider the solutions of the CYM to the leading order in the weak field.
Owing to the equations of motions we get
\begin{eqnarray}
	j^{(1)} &=& 0, \\
	j^{(1)}_\perp &=&  0\,.
\end{eqnarray}
Because of the absence of the currents, there are no  non-zero contributions from the upper light-cone.
The  surface term for the transverse component is defined by the initial conditions
\begin{align}
	\mathfrak{S}^{(1)}_\perp(\tau,\vp{k})
	 &= -
	\lim_{\tau \to 0+} \left(
	\tau \partial_\tau H_0^{(1)} (k_\perp \tau)
	\beta^{(1)}_\perp (\tau, \vp{k}) \right) \notag \\&=
	-
	\frac{2}{\pi} i
	\beta^{(1)}_\perp(\tau=0, \vp{k}) =
	\frac{2 i }{\pi k_\perp} b_2(\vp{k})
	\,.
\end{align}
Correspondingly, the contribution from the $\eta$ component is given by
\begin{equation}
	\mathfrak{S}^{(1)}_\eta(\tau,\vp{k}) =
-  \frac{4}{\pi} i \frac{\beta(\tau=0,\vp{k})}{k_\perp} = -
	\frac{2 i }{\pi k_\perp} b_1(\vp{k})\,.
\end{equation}

Combining these equations together,
we conclude that the single inclusive gluon distribution to this order is given by
\begin{align}
\label{Eq:leadingorder}
	E_k \frac{d N}{d^3 k}  =
	\frac{1}{2\pi^3}
	&\left[
		\beta_i(\tau=0, \vp{k})
		\mathfrak{t}_{ij} (\vp{k})
		\beta_j(\tau=0, -\vp{k})
	\right. \notag \\ &\left.
		+ \frac{4}{k^2_\perp}
	\beta(\tau=0, \vp{k})
	\beta(\tau=0, -\vp{k})
	\right]\;,
\end{align}
where $	\mathfrak{t}_{ij} (\vp{k})$ is  the two-dimensional transverse projector
\begin{equation}
	\mathfrak{t}_{ij} (\vp{k}) = \delta_{ij} - \frac{k_i k_j }{k_\perp^2}.
\end{equation}

This expression is manifestly symmetric under $\vp{k}\to-\vp{k}$.
To match this result to the one derived previously~\cite{Dumitru:2001ux}, we rewrite
Eq.~\eqref{Eq:leadingorder}
in the form~\footnote{See Eq.~\eqref{Eq:t_to_e}.}
\begin{eqnarray}
	E_k \frac{d N}{d^3 k}  &=&
	\frac{1}{2\pi^2}
 {\rm tr}
  \left( |a_1|^2 + |a_2|^2 \right)
	\label{Eq:dNdk1}
\end{eqnarray}
where
\begin{eqnarray}
	a_1 &=& \frac{g}{\sqrt{\pi} k_\perp}
	\int d^2 x_\perp
	e^{-i \vp{k} \vp{x}}
	U^\d(\vp{x})
	[
		\alpha^{(1) i}_1(\vp{x}) ,
		\alpha_2^i(\vp{x})
	]
	U(\vp{x}),\\
	a_2 &=& \frac{1}{\sqrt{\pi} k_\perp}
	\int d^2 x_\perp
	e^{-i \vp{k} \vp{x}}
	\epsilon_{ij} \partial_j
	U^\d(\vp{x})
	\alpha^{(1) i}_1(\vp{x})
	U(\vp{x})\,.
\end{eqnarray}
This  coincides with Dumitru--McLerran result~\cite{Dumitru:2001ux} modulo an
irrelevant complex phase in the definition of $a_{1,2}$.

Further simplifications are possible if explicitly expand
the weak field into a power series in $\rho_1$~\footnote{See Eq.~\eqref{Eq:U1_expansion}.}
\begin{equation}
	\alpha^{(1) i}_1 (\vp{x}) = \partial^i \Phi_1(\vp{x}) =
	g \frac{\partial^i}{\partial_\perp^2} \rho_1(\vp{x}).
\end{equation}
Substituting to Eq.~\eqref{Eq:dNdk1} we obtain
\begin{align}
	E_k \frac{d N}{d^3 k}  &=
	\frac{1}{4\pi^3 k_\perp^2}
  (
	\delta_{ij} \delta_{lm} +
	\epsilon_{ij} \epsilon_{lm}
)
\Omega_{ij}^b(\vp{k})
\left[\Omega_{lm}^b(\vp{k})\right]^*
\notag \\ &=
	\frac{g^2}{4\pi^3 k_\perp^2}
  (
	\delta_{ij} \delta_{lm} +
	\epsilon_{ij} \epsilon_{lm}
	) \notag \\
	&\times
	\int
	\frac{d^2 p_\perp}{(2\pi)^2}
	\frac{d^2 q_\perp}{(2\pi)^2}
	\frac{p_{\perp, i} (k-p)_{\perp, j} }{p_\perp^2}
	\frac{q_{\perp, l} (k-q)_{\perp, m} }{q_\perp^2}
	\notag \\
	&\quad\quad\quad \times
	\rho_a^*(\vp{q})
	\left[
	W^\dagger(\vp{k}-\vp{q})
	W(\vp{k}-\vp{p})
	\right]_{ab}
	\rho_b(\vp{p}) ,
	\label{Eq:Gelis_form}
\end{align}
where we introduced the Fourier transforms of
the components of Eq.~\eqref{Eq:A}
\begin{equation}
\Omega_{ij}^b(\vp{k}) =
	g \int \frac{d^2 p_\perp}{(2\pi)^2}
	\frac{p_{\perp, i} (k-p)_{\perp, j} }{p_\perp^2}
	\rho_a(\vp{p}) W_{ba} (\vp{k}-\vp{p})
\label{Eq:OmegaDef}
\end{equation}
to simplify the notation in the coming section.
In Appendix A, we provide yet another alternative form of Eq.~\eqref{Eq:dNdk1}.

\section{Second order}
At second order we expect some non-trivial modification of the
particle production owing to  the presence of
non-trivial currents
\begin{eqnarray}
	j^{(2)} (\tau, \vp{x})  &=& -ig \left(
	\partial_i [\beta^{(1)}_i(\tau,\vp{x}), \beta^{(1)}(\tau, \vp{x})   ]\right.
	\notag
	\\ &&\left. +
	[\beta^{(1)}_i(\tau,\vp{x}), 	\partial_i \beta^{(1)}(\tau, \vp{x})   ]
	\right), \\
	\label{Eq:j2}
	j^{(2)}_i  (\tau, \vp{x})  &=&
	- \partial_i \partial_j \beta^{(2)}_j (\tau, \vp{x} ) - ig \left(
	\partial_j [\beta^{(1)}_j(\tau,\vp{x}), \beta^{(1)}_i(\tau, \vp{x})   ]
	\right. \\ && \left.
	+
	[\beta^{(1)}_j(\tau,\vp{x}), 	\partial_j \beta^{(1)}_i(\tau, \vp{x})
	-\partial_i \beta^{(1)}_j (\tau, \vp{x})
	] \right. \notag \\ \notag
	&&\left.
	-\tau^2 [\beta^{(1)}(\tau,\vp{x}), \partial_i \beta^{(1)}(\tau, \vp{x})]
	\right)\notag .
\end{eqnarray}
Note that to this order, we do not have to solve the equations of motion for $\beta^{(2)}$.
The
contribution of the currents to particle production  is solely defined by
combinations of $\beta^{(1)}$ except for the
term proportional to the gradient of the divergence of $\beta^{(2)}$
(the first term in Eq.~\eqref{Eq:j2}).
Fortunately this term does not contribute to the transverse current $j_\perp$
and thus drops out from the particle production equations,
see Eq.~\eqref{Eq:Bulk_perp}  and Eq.~\eqref{Eq:j_perp_k}.
This becomes  obvious in  momentum space~\footnote{The two-dimensional cross-product is defined as $\vp{a} \times \vp{b} = \epsilon_{ij} a_i b_j$.}
\begin{align}
	\notag
	j^{(2)} (\tau, \vp{k})  &= g \int \frac{d^2q}{(2\pi)^2}   [ (2 \vp{k} -\vp{q}) \cdot  \vec{\beta}^{(1)} (\tau, \vp{q}) , \beta^{(1)}  (\tau, \vp{k} - \vp{q}) ], \\
	j^{(2)}_\perp (\tau, \vp{k}) &=
	g
	\int \frac{d^2q}{(2\pi)^2}
	\left(
	i \left[ (2 \vp{k} - \vp{q}) \cdot \vec{\beta} ^{(1)} (\tau, \vp{q}) , \frac{\vec{\beta} ^{(1)} (\tau, \vp{k} - \vp{q}) \times \vp{k}}{k_\perp} \right]
	+ \right.
	\notag
	\\&
	\quad \quad \quad \quad \left.
	i\, \frac{\vp{q}\times\vp{k}}{k_\perp} \left( \tau^2  [  \beta ^{(1)} (\tau, \vp{q}) , \, \beta ^{(1)} (\tau, \vp{k} - \vp{q}) ] \right. \right.
	\notag \\
	&\quad \quad \quad \quad  \left. \left.
	+  [ \vec{\beta} ^{(1)} (\tau, \vp{q}),\,  \vec{\beta} ^{(1)}  (\tau, \vp{k} - \vp{q})]
	\right)
	\right). \notag
\end{align}

\subsection{ $\eta$-component of  bulk contribution  }
The goal of this subsection is to compute
\begin{equation}
	\mathfrak{B}^{(2)} _\eta(\vp{k}) = \int d \tau \tau^2
	H_1^{(1)} (k_\perp \tau)
	j^{(2)} (\tau, \vp{k})\,.
\end{equation}
For this we note two useful identities obtained based on the equations from
Appendix B:
\begin{eqnarray}
	&&\int d\tau \tau H^{(1)}_1 (k_\perp \tau)
	J_0(q_\perp \tau)
	J_1(|\vp{k}-\vp{q}|\tau) =
	\frac{1}{\pi} \frac{1}{|\vp{k}-\vp{q}| k_\perp} \times \nonumber \\
&&	\left( \frac{(\vp{k}-\vp{q})\cdot \vp{k}}{|\vp{q}\times\vp{k}|} - i  \right)\,,\\
	&& \int d\tau \tau H^{(1)}_1 (k_\perp \tau)
	J_1(|\vp{k}-\vp{q}|\tau) =
	i \frac{2}{\pi} \frac{|\vp{k}-\vp{q}|}{k_\perp} \frac{1}{k_\perp^2 - |\vp{k}-\vp{q}|^2}\,.
\end{eqnarray}
Thus the bulk contribution for the $\eta$-component at second order reads
\begin{eqnarray}
	\mathfrak{B}^{(2)}_\eta(\vp{k})  &=&   \frac{2 ig}{\pi k_\perp}   \int \frac{d^2q}{(2\pi)^2}
	\left(
	\frac{\vp{k}\times\vp{q}}{q^2|\vp{k}-\vp{q}|^2} \left(\frac{(\vp{k}-\vp{q})\cdot \vp{k}}{|\vp{q}\times\vp{k}|} - i    \right) \right. \times \nonumber \\
 & &	\left.  [b_2(\vp{q}),\, b_1(\vp{k}-\vp{q})]
	\notag  + i [\Lambda(\vp{q}),\, b_1(\vp{k}-\vp{q})]
	\right) .
\end{eqnarray}

\subsection{ Transverse vector-component of  bulk contribution}
Analogously, using the integrals from Appendix B we get
\begin{align}
	\mathfrak{B}^{(2)} _\perp (\vp{k}) &= \int d \tau \tau
	H_0^{(1)} (k_\perp \tau)
	j^{(2)}_\perp (\tau, \vp{k}) \notag \\
	&= \notag g \frac{2}{\pi k_\perp}
	\int \frac{d^2q}{(2\pi)^2}
	\left(
	\frac{1}{2}
	\vp{k} \times  \vp{q} [\Lambda(\vp{q}), \Lambda(\vp{k}-\vp{q})]
	\right. \notag \\& \notag
	\left.
	- \frac{\vp{k} \cdot \vp{q}}{q_\perp^2}  [b_2(\vp{q}), \Lambda(\vp{k}-\vp{q})]
	\right. \\ \notag
	&- i \frac{1}{2} \frac{ \vp{k} \times \vp{q}}{ | \vp{k} \times \vp{q}| }
	\frac{k_\perp^2 + \vp{q} \cdot (\vp{q} - \vp{k})}{q_\perp^2 |\vp{k}-\vp{q}|^2}
	 [b_2(\vp{q}), b_2(\vp{k}-\vp{q})]
	 \\
	 &\left.
	 + \frac{1}{2} \frac{\vp{k}\times\vp{q}}{q_\perp^2  |\vp{k}-\vp{q}|^2}
	 \left(
	 1 + i \frac{\vp{q} \cdot (\vp{k}-\vp{q})}{| \vp{k} \times \vp{q}|}
	 \right)
	 [b_1(\vp{q}), b_1(\vp{k}-\vp{q})]
	\right)\,.
\end{align}
Although the last equation is complicated, we expect significant simplifications
for the asymmetric part, as we will demonstrate below.

\subsection{Surface contributions}
To obtain the final equation for particle production we have to derive the
surface contributions as well.
They are
\begin{equation}
\mathfrak{S}^{(2)}_\eta (\vec{k})
= - i \frac{4}{\pi} \frac{\beta^{(2)} (\tau=0,\vp{k})}{k_\perp}
\end{equation}
and
\begin{equation}
\mathfrak{S}^{(2)}_\perp (\vec{k})
	 =
	-  i
	\frac{2}{\pi}
	\beta^{(2)}_\perp(\tau=0, \vp{k})\,,
\end{equation}
where the functions are defined by the  second-order
expansion coefficient in the weak field of the initial conditions
\begin{eqnarray}
	\beta^{(2)} (\tau\to 0, \vp{x}) &=& \frac{ig}{2}
	U^\d(\vp{x})
	[\alpha_1^{(2) \, i}(\vp{x}) ,\, \alpha_2^{i}(\vp{x})]
	U(\vp{x}), \\
	\beta^{(2)}_i (\tau\to 0, \vp{x}) &=&
	U^\d(\vp{x})
	\alpha_1^{(2) \, i}(\vp{x})
	U(\vp{x}).
\end{eqnarray}
Here the weak proton field at second order is given by~\footnote{See Eq.~\eqref{Eq:U1_expansion}.}
\begin{equation}
	\alpha_1^{(2) \,i} =
	\left(
	\delta_{ij} - \frac{\partial_i \partial_j}{\partial^2}
	\right) \left[ \partial^j \Phi_1 (\vp{x}), \Phi_1 (\vp{x})   \right]\;.
\end{equation}

\begin{figure}
	\centering
	\includegraphics[width=\textwidth]{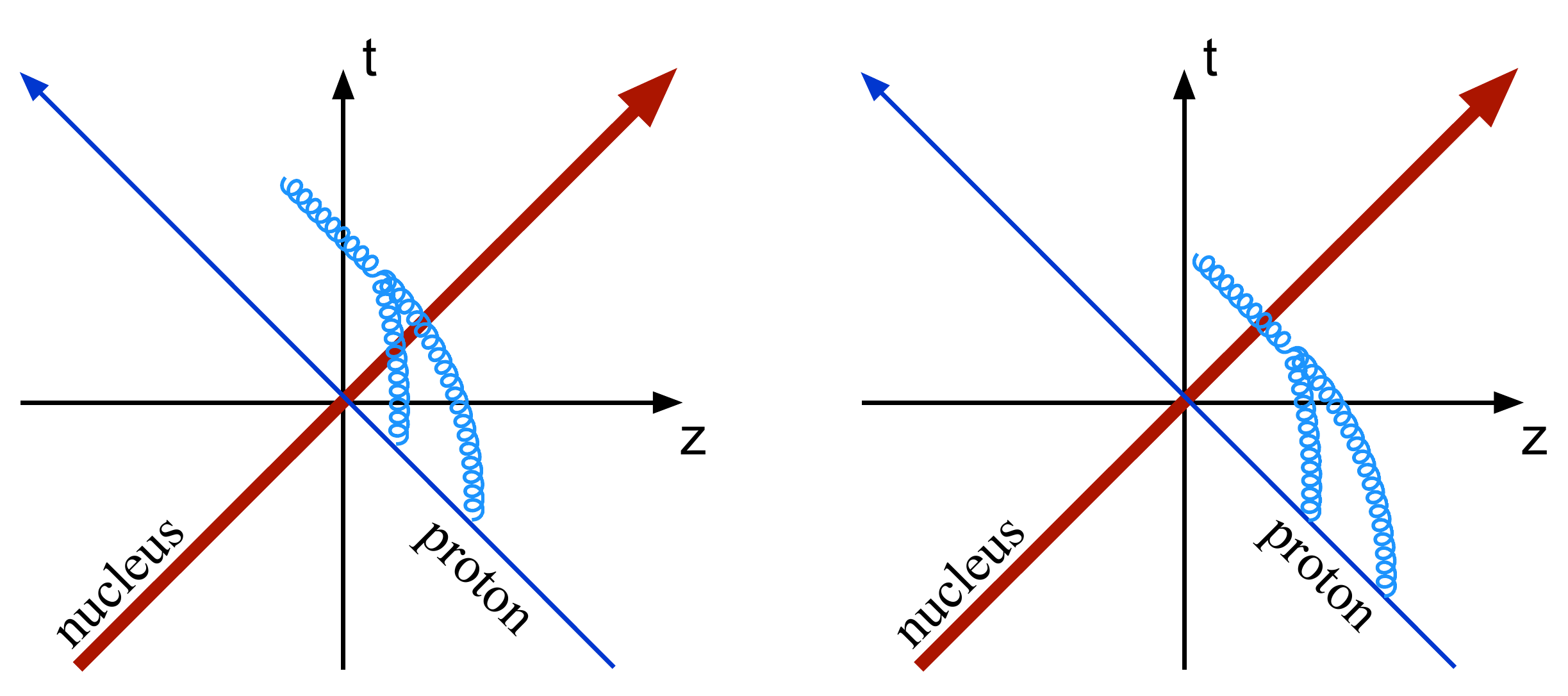}
	\caption{Illustration of the bulk (left) and the surface (right) contributions to the amplitude at order $g^3$.}
	\label{fig:bulksurf}
\end{figure}

\subsection{Odd azimuthal anisotropy on event-by-event basis}
The goal of this section is to show that the single inclusive
particle production configuration-by-configuration (before performing the average
with respect to $\rho_1$ and $\rho_2$)
has odd azimuthal harmonics at
second order. This can be straight-forwardly shown using  the
results obtained in the previous sections. We however prefer to
use the following line of argumentation. Lets consider the
single inclusive cross-section
\begin{align}
	E_k \frac{d N}{d^3 k}  &=
	\sum_{\gamma = \eta, _\perp}
	(a^{(1)}_\gamma(\vp{k}) + a^{(2)}_\gamma(\vp{k}))
	(a^{(1)}_\gamma(\vp{k}) + a^{(2)}_\gamma(\vp{k}))^*.
\end{align}
As we discussed previously the first order is entirely defined by the
surface contribution, i.e.  $ a^{(1)}_\gamma(\vp{k}) = \mathfrak{S}^{(1)}_\gamma(\vp{k}) $
with the following property for $   \mathfrak{S}^{(1)}_\gamma (\vp{k})$
\begin{align}
	 \mathfrak{S}^{(1)}_\gamma (\vp{k})   = - ( \mathfrak{S}^{(1)}_\gamma  (-\vp{k})  )^* \;.
\end{align}
An analogous relation holds also for  second order
\begin{align}
	 \mathfrak{S}^{(2)}_\gamma (\vp{k})   = - ( \mathfrak{S}^{(2)}_\gamma  (-\vp{k})  )^* \;.
\end{align}

The asymmetric part of the single inclusive production is
\begin{align}
	&\frac{E_k}{ 2} \left( \frac{d N (\vp{k})}{d^3 k}
	-	\frac{d N (-\vp{k})}{d^3 k} \right) \\ & =
	\frac12 \sum_{\gamma = \eta, _\perp}
	(a^{(1)}_\gamma(\vp{k}) + a^{(2)}_\gamma(\vp{k}))
	(a^{(1)}_\gamma(\vp{k}) + a^{(2)}_\gamma(\vp{k}))^*  \notag \\ &
-	\frac12 \sum_{\gamma = \eta, _\perp}
	(a^{(1)}_\gamma(-\vp{k}) + a^{(2)}_\gamma(-\vp{k}))
	(a^{(1)}_\gamma(-\vp{k}) + a^{(2)}_\gamma(-\vp{k}))^*
	\notag \\
	& =
	{
  \frac{1}{8\pi} }
	\Re \left( (\mathfrak{S}^{(1)}_\gamma (\vp{k}))^* \left[
		\mathfrak{B}^{(2)}_\gamma (\vp{k}) +
		(\mathfrak{B}^{(2)}_\gamma (-\vp{k}))^*
 		+ \mathfrak{S}^{(2)}_\gamma (\vp{k})
 		+ (\mathfrak{S}^{(2)}_\gamma (\vp{k}))^*
	\right]\notag
\right)\\ &\notag  + {\cal O} (\rho_1^4)
=
	{
  \frac{1}{8\pi} }
\Re \left( (\mathfrak{S}^{(1)}_\gamma (\vp{k}))^* \left[
		\mathfrak{B}^{(2)}_\gamma (\vp{k}) +
		(\mathfrak{B}^{(2)}_\gamma (-\vp{k}))^*
	\right]
\right) + {\cal O} (\rho_1^4) .
	\label{Eq:Odd_ass}
\end{align}
The surface contribution in the square bracket cancels, as we alluded to before.
In order to compute the bulk contribution in the square brackets,
lets go back and consider Eq.~\eqref{Eq:Bulk}.
Since the functions $\beta(\vp{x})$  and $\beta_\perp(\vp{x})$ are real we have
\begin{align}
		\mathfrak{B}^{(2)}_\eta (\vp{k}) +
		(\mathfrak{B}^{(2)}_\eta (-\vp{k}))^*
	& =
	2 \int_0^\infty d \tau \tau^2
	J_1 (k_\perp \tau)
	j^{(2)} (\tau, \vp{k})
\end{align}
and
\begin{align}
		\mathfrak{B}^{(2)}_\perp (\vp{k}) +
		(\mathfrak{B}^{(2)}_\perp (-\vp{k}))^*
	& =
	2 \int_0^\infty d \tau \tau
	J_0 (k_\perp \tau)
	j^{(2)}_\perp (\tau, \vp{k})\;.
\end{align}
The cancellation of the Neumann functions
simplifies the computation of the right-hand side
\begin{align}
	&\mathfrak{B}^{(2)}_\eta (\vp{k}) +
	(\mathfrak{B}^{(2)}_\eta (-\vp{k}))^*
	\notag
	\\
	&= \frac{4 g}{k_\perp \pi}
	\int \frac{d^2 q}{(2\pi)^2}
	\frac{\vp{k}\times \vp{q}}{|\vp{k}\times \vp{q}|}
	\frac{(\vp{k}-\vp{q})\cdot\vp{k}}{q_\perp^2 |\vp{k}-\vp{q}|^2 }
	i [b_{ 2} (\vp{q}), b_{1}(\vp{k}-\vp{q})]
	\label{Eq:Odd_Beta}
\end{align}
and
\begin{align}
	\label{Eq:Odd_Bperp}
	&\mathfrak{B}^{(2)}_\perp (\vp{k}) +
	(\mathfrak{B}^{(2)}_\perp (-\vp{k}))^*
	=
	\notag \\
	&\quad - \frac{2 g}{k_\perp \pi}
	\int \frac{d^2 q}{(2\pi)^2}
	\frac{\vp{k}\times \vp{q}}{|\vp{k}\times \vp{q}|}
	\frac{1}{q_\perp^2 |\vp{k}-\vp{q}|^2 }
	\notag \\&\quad
	\left(
	(k_\perp^2 + \vp{q}\cdot(\vp{q}-\vp{k}))
	i [b_2(\vp{q}), b_2(\vp{k}-\vp{q})]
	\right. \notag
	\\
	&\quad
	\left.
	-\vp{q}\cdot(\vp{k}-\vp{q})
	i [b_1(\vp{q}), b_1(\vp{k}-\vp{q})]
	\right) \,.
\end{align}
It is remarkable that the gauge field $\Lambda(\vp{k})$ does not contribute to  this expression.
Both expressions in Eqs.
\eqref{Eq:Odd_Beta}
and
\eqref{Eq:Odd_Bperp}
are non-local owing to the presence of the
ratio
\begin{equation}
	\frac{\vp{k}\times \vp{q}}{|\vp{k}\times \vp{q}|}
	= \rm {sign} \left[ \; \sin(\phi_{\angle(\vp{k}, \vp{q})}) \right]\;.
\end{equation}

Summing everything up, the odd contribution is given by the following
\begin{align}
	&\frac{E_k}{ 2} \left( \frac{d N (\vp{k})}{d^3 k}
	-	\frac{d N (-\vp{k})}{d^3 k} \right) = \notag \\
	&\quad
	{\frac{1}{8\pi} }
	\Re \left(
	 \frac{4 ig}{\pi^2 k_\perp^2} b^\star_2(\vp{k})
	\int \frac{d^2 q}{(2\pi)^2}
	\frac{\vp{k}\times \vp{q}}{|\vp{k}\times \vp{q}|}
	\frac{1}{q_\perp^2 |\vp{k}-\vp{q}|^2 }
	\right.\notag  \\ & \quad \left.
	\left(
	(k_\perp^2 + \vp{q}\cdot(\vp{q}-\vp{k}))
	i [b_2(\vp{q}), b_2(\vp{k}-\vp{q})]
	\right. \right. \notag
	\\
	&\quad
	\left. \left.
	-\vp{q}\cdot(\vp{k}-\vp{q})
	i [b_1(\vp{q}), b_1(\vp{k}-\vp{q})]
	\right) +
	\right.
	\notag
	\\
	& \quad \quad \left.+ \frac{8 ig}{\pi^2 k_\perp^2} b^\star_1(\vp{k})
	\int \frac{d^2 q}{(2\pi)^2}
	\frac{\vp{k}\times \vp{q}}{|\vp{k}\times \vp{q}|}
	\frac{(\vp{k}-\vp{q})\cdot\vp{k}}{q_\perp^2 |\vp{k}-\vp{q}|^2 }
	i [b_{ 2}(\vp{q}), b_{1}(\vp{k}-\vp{q})]
	\right) \notag = \\
    &=
	{ \frac{1}{8\pi} }
	\Im
	\left\{
		\frac{2g}{\pi^2 k_\perp^2}
		\int \frac{d^2 q}{(2\pi)^2}
		\frac{\vp{k}\times \vp{q}}{|\vp{k}\times \vp{q}|}
		\frac{1}{q_\perp^2 |\vp{k}-\vp{q}|^2 }
		\right. \notag \\
		&\quad \left.
		\times
		f^{abc}
		\Omega^a_{ij} (\vp{q})
		\Omega^b_{mn} (\vp{k}-\vp{q})
		\Omega^{c\star}_{rp} (\vp{k})
		\notag
		\times \right.  \\ & \quad
		\left.
		\left[
			\left(
			k_\perp^2 \epsilon^{ij} \epsilon^{mn}
		-\vp{q} \cdot (\vp{k} - \vp{q} )
		(\epsilon^{ij} \epsilon^{mn}+\delta^{ij} \delta^{mn})
		\right) \epsilon^{rp}
		\right. \right.\notag \\ & \quad \left. \left.
		+
		2 \vp{k} \cdot (\vp{k}-\vp{q}) {\epsilon^{ij} \delta^{mn}} \delta^{rp}
		\right]
	\right\} ,
	\label{Eq:FinalAssymitry}
\end{align}
where $\Omega$ is defined in Eq.~\eqref{Eq:OmegaDef}.

\section{Double inclusive gluon production in  leading log}
\label{Sec:DIP}
The double inclusive gluon production at the leading log approximation reads~\cite{Gelis:2008ad}
\begin{equation}
	E_k E_q \frac{d\; \overline{N}}{d^3k d^3 q}  =
	\left\langle
	E_k \frac{d N}{d^3 k }
	E_q \frac{d N}{d^3 q}
	\right\rangle\;,
\end{equation}
where the average is performed over the target and the projectile fields.
In a Gaussian ensemble,  the average removes all contributions odd in $\rho_1$.

To simplify the notation we define
\begin{equation}
	E_k \frac{d N}{d^3 k }  = n^{(2)}(\vp{k}) +  n^{(3)}(\vp{k})  +  n^{(4)}(\vp{k}) + \dots\;,
\end{equation}
where according to previously used definitions
\begin{align}
	n^{(2)} (\vp{k}) &= \sum_{\gamma=\eta,\perp} |a^{(1)}_\gamma(\vp{k})|^2 \;, \\
	n^{(3)} (\vp{k}) &= \sum_{\gamma=\eta,\perp}
	a^{(1)}_\gamma(\vp{k}) \left( a^{(2)}_\gamma(\vp{k}) \right)^* + {\rm c.c.}  \;, \\
	n^{(4)} (\vp{k}) &= \sum_{\gamma=\eta,\perp} \left[
	a^{(1)}_\gamma(\vp{k}) \left( a^{(3)}_\gamma(\vp{k}) \right)^* +
	a^{(3)}_\gamma(\vp{k}) \left( a^{(1)}_\gamma(\vp{k}) \right)^*
+  |a^{(2)}_\gamma(\vp{k})|^2 \right]
	\;.
\end{align}
As we established earlier, to the leading order the cross-section is symmetric
configuration-by-configuration
\begin{equation}
	n^{(2)} (\vp{k}) = n^{(2)} (- \vp{k}) .
\end{equation}
In addition, the condition that
\begin{equation}
	E_k E_q \frac{d\; \overline{N}}{d^3k d^3 p} (\vp{k}, \vp{p})
	=
	E_k E_q \frac{d\; \overline{N}}{d^3k d^3 p} (-\vp{k}, -\vp{p})
\end{equation}
leads to
\begin{equation}
	\left\langle \left( n_\gamma^{(4)}(\vp{k}) - n_\gamma^{(4)}(-\vp{k})  \right)
	n^{(2)}_\gamma(\vp{p}) 
	\right\rangle = 0 \;.
\end{equation}
This guarantees  that the contribution to the odd asymmetry
depends only on $n^{(3)}$ computed in the previous section.
Indeed
\begin{equation}
	\frac{E_k E_q}{2} \left(  \frac{d \overline{N}}{d^3 k d^3 q  }   (\vp{k}, \vp{p})   -
	\frac{d \overline{N}}{d^3 k d^3 p }   (-\vp{k}, \vp{p})
\right) = \frac12
\left\langle
\left(
n^{(3)}_\gamma(\vp{k})
-
n^{(3)}_\gamma(- \vp{k})
\right)
n^{(3)}_\gamma(\vp{p})
\right\rangle.
\end{equation}
In this notation, the difference
$\frac12\left(
n^{(3)}_\gamma(\vp{k})
-
n^{(3)}_\gamma(- \vp{k})
\right)
$ is given entirely  by Eq.~\eqref{Eq:FinalAssymitry}.
This contribution is non-vanishing and gives rise to
odd azimuthal anisotropy. It is obviously connected to
the initial state distribution of the color charges, but
has some non-local dependence on spatial points.
Most importantly this contribution
comes from the evolution of the field in the forward light cone and is not just defined
by the initial conditions on the light-cone as at the leading order.

To proceed further
we will consider an
expression asymmetrized both with respect to $\vp{k}$ and
$\vp{q}$:
\begin{equation}
\frac14
\left(
n^{(3)}_\gamma(\vp{k})
-
n^{(3)}_\gamma(- \vp{k})
\right)
\left(
n^{(3)}_\gamma(\vp{p})
-
n^{(3)}_\gamma(- \vp{p})
\right).
\label{Rq:n3}
\end{equation}
As we established in the previous section, each difference is
proportional to the imaginary part of some function $f$, i.e
\begin{align*}&\frac{1}{2} \left(
n^{(3)}_\gamma(\vp{p})
-
n^{(3)}_\gamma(- \vp{p})
\right) = \Im f(\vp{p})
=
\frac{1}{2 i } \left( f(\vp{p}) - f^*(\vp{p}) \right) \\
&
=
\frac{1}{2 i } \left( f(\vp{p}) - f(-\vp{p}) \right) .
\end{align*}
Thus for our purpose, we can just equate
$ f(\vp{p})   = i n^{(3)}_\gamma(\vp{p}) $. This assumptions is
not true
in general but is sufficient for the current calculations of the
asymmetric part
\begin{align}
&\frac14
\left(
n^{(3)}_\gamma(\vp{k})
-
n^{(3)}_\gamma(- \vp{k})
\right)
\left(
n^{(3)}_\gamma(\vp{p})
-
n^{(3)}_\gamma(- \vp{p})
\right)
 =  \notag \\ &
 - \frac{1}{4}
 \left(
 f(\vp{p}) - f(-\vp{p})
 \right)
 \left(
 f(\vp{k}) - f(-\vp{k})
 \right)
= \notag \\  &
- \frac{1}{4}
\left( \left[ f(\vp{p}) f(\vp{k}) - (\vp{k}\to-\vp{k}) \right]
- (\vp{p}\to-\vp{p})
\right).
\end{align}
Therefore it is enough to consider only one term, e.g. $ f(\vp{p}) f(\vp{k})$;
the rest of the  terms can be obtained by changing the direction of momenta.
In Appendix C we derived the expression for  $\langle
f(\vp{p}) f(\vp{k}) \rangle_{\rho_1}$.
In has fifteen different terms and must be further averaged with respect to the target field.
This would generate over 125 terms only for $\langle
f(\vp{p}) f(\vp{k}) \rangle_{\rho_1, \rho_2}$. At this point we see that the only reasonable resolution
would be to perform numerical simulations where averages with respect to the projectile and target configurations are performed using Monte-Carlo technique. We postpone this for further publications.

\section{Summary and conclusions}
Here we briefly summarize our results  and provide some comments.
\begin{enumerate}
 \item{The surface contribution on the light cone gives zero odd azimuthal anisotropy to
all orders.
It is T-even and can be written in a local form.}

\item{The odd harmonics originate from evolution in the forward light cone. They are non-local
and not T-even. In single particle inclusive process they average out to zero
for a Gaussian ensemble because they are proportional to $\rho_1^3$. Essentially they are defined by
odderon exchanges.}

\item{We were unable to establish the connection between our
		formulae and  geometric  anisotropy in the initial
state $\epsilon_3$. From the equation it is obvious that
the anisotropy is not defined by the global scales, but rather by the
geometry on the scales of  $1/Q_s$.}

 \item{The argument presented in Ref.~\cite{Kovner:2010xk,Kovner:2012jm} is valid only for the surface
contribution in the dilute approximation. We showed that the bulk contribution for
configuration-by-configuration single inclusive result  is not symmetric under
$\vp{k} \to -\vp{k}$.}

\item{Our results take into account the first saturation correction, which was
also considered in~Ref.~\cite{Chirilli:2015tea}.}

\item{We complement the numerical results of Refs.~\cite{Lappi:2009xa,Schenke:2015aqa}
with an  analytical prove beyond any doubts in numerics and uncertainty in the
prescription of what is defined by a gluon at an intermediate state, $\tau$, that
CYM produces odd azimuthal anisotropy.}

\item{Our results  can be potentially used to calculate $v_3$ without solving CYM
numerically.}

\end{enumerate}

\section{Acknowledgements}
We thank A.~Bzdak for sharing his puzzle on the two-dimensional Fourier transformation, odd harmonics, and necessity to include the time-dependence.
We thank A. Dumitru,  M. Sievert, H.-U. Yee,    and especially  A. Kovner, M. Lublinsky
and R. Venugopalan
for useful discussions.   L. McLerran  was  supported under Department of Energy contract number Contract No. DE-SC0012704 at Brookhaven National Laboratory,
and grant number grant No. DE-FG02-00ER41132 at Institute for Nuclear Theory.

\section{Note added after publication} 
The results  obtaind here in Fock-Schwinger gauge $A_\tau =0$ 
were reproduced and  extedned in the global $A^+=0$ gauge in Ref.~\cite{Kovchegov:2018jun}. 
Phenomenological calculations were performed in Ref.~\cite{Mace:2018vwq}. 
The effect of quantum evolution in the projectile was considered in Ref.~\cite{Kovner:2016jfp}.

\section{Appendix A: Leading order results in coordinate space} \label{appA}
Equation~\eqref{Eq:Gelis_form} can be rewritten in an alternative form.
Lets consider the combination
\begin{align}
	&\frac{\delta_{ij} \delta_{lm} + \epsilon_{ij} \epsilon_{lm}}{k_\perp^2}
	\frac{p_{\perp,i} ( k-p)_{\perp,j}}{p_\perp^2}
	\frac{q_{\perp,l} ( k-q)_{\perp,m}}{q_\perp^2}
	\notag \\ &=
	\frac{\vp{p}\cdot(\vp{k}-\vp{p})\  \vp{q}\cdot(\vp{k}-\vp{q}) + (\vp{p}\times \vp{k}) \   (\vp{q}\times \vp{k}) } {k_\perp^2 p_\perp^2 q^2_\perp} .
\end{align}
The last expression can be further simplified using the identity
\begin{equation}
	(\vp{p}\times \vp{k}) \   (\vp{q}\times \vp{k}) = (\vp{p}\cdot \vp{q})  \  k_\perp^2
	- (\vp{p} \cdot \vp{k}) \ (\vp{q} \cdot \vp{k} )
\end{equation}
which can be proven starting from the  identity
\begin{equation}
    (\vp{k}\times \vp{u})^2 = k_\perp^2 u_\perp^2 -  (\vp{k}\cdot \vp{u})^2
\end{equation}
and proceeding by substituting $\vp{u} = \vp{p}+\vp{q}$.
Thus
\begin{align}
	&\frac{\delta_{ij} \delta_{lm} + \epsilon_{ij} \epsilon_{lm}}{k_\perp^2}
	\frac{p_{\perp,i} ( k-p)_{\perp,j}}{p_\perp^2}
	\frac{q_{\perp,l} ( k-q)_{\perp,m}}{q_\perp^2}
	\notag \\ &=
	\left(
	\frac{\vp{k}}{k_\perp^2}
	-
	\frac{\vp{p}}{p_\perp^2}
	\right)
	\cdot
	\left(
	\frac{\vp{k}}{k_\perp^2}
	-
	\frac{\vp{q}}{q_\perp^2}
	\right) .
\end{align}
Substituting this into Eq.~\eqref{Eq:Gelis_form} we get
\begin{align}
	E_k \frac{d N}{d^3 k}  &=
	\frac{2g^2}{(2\pi)^3}
	\int
	\frac{d^2 p_\perp}{(2\pi)^2}
	\frac{d^2 q_\perp}{(2\pi)^2}
		\left(
	\frac{\vp{k}}{k_\perp^2}
	-
	\frac{\vp{p}}{p_\perp^2}
	\right)
	\cdot
	\left(
	\frac{\vp{k}}{k_\perp^2}
	-
	\frac{\vp{q}}{q_\perp^2}
	\right)
	\notag \\ &\times
	\rho_a^*(\vp{p})
	\left[W^\dagger(\vp{k}-\vp{p})
	W(\vp{k}-\vp{q}) \right]^{ab}
	\rho_b(\vp{q})
	\label{Eq:Kovner_form}
\end{align}
or in coordinate space
\begin{align}
	E_k \frac{d N}{d^3 k}  &= \frac{2 \alpha_s}{\pi}
	\int_{u} \int_{v} \int_{x} \int_{y}
	e^{i \vp{k} (\vp{u}-\vp{v})}
	\frac{\vec{v}-\vec{y}}{|\vec{v}-\vec{y}|^2}
	\frac{\vec{x}-\vec{u}}{|\vec{v}-\vec{x}|^2} \times
	\\
	& \rho^a (\vp{x})
	\left(
	\left[ W^\dagger (\vp{x})  - W^\dagger(\vp{u})
	\right]
	\left[ W (\vp{y})  - W(\vp{v})
	\right]
	\right)^{ab}
	\rho^b (\vp{y})\;.
\end{align}


\section{Appendix B:  List of useful integrals and relations} \label{appB}
Here we collect the list of useful integrals and relations.
Some integrals are adopted from more general ones of Ref.~\cite{prudnikov_integrals_1998}
\begin{eqnarray}
	&&\int d \tau \tau J_\nu (p_\perp \tau) J_\nu(k_\perp \tau)
	= \frac{\delta(p_\perp-k_\perp)}{k_\perp}\;,\\
	&&\int d \tau \tau J_0(p_\perp \tau) Y_0(k_\perp \tau)
	= \frac{2}{\pi} \frac{1}{k_\perp^2-p_\perp^2}\;, \\
	&&\int d \tau \tau J_1(p_\perp \tau) Y_1(k_\perp \tau)
	= \frac{2}{\pi} \frac{p_\perp}{k_\perp} \frac{1}{k_\perp^2-p_\perp^2}\;, \\
	&&\int d \tau \tau H^{(1)}_0(k_\perp \tau) J_0(|\vec{q}_\perp-\vec{k}_\perp| \tau) J_0(q_\perp \tau)
	=  \frac{1}{\pi} \frac{1}{|\vp{k}\times\vp{q}|}\;,\\
	&&\int d \tau \tau J_1(q_\perp \tau) J_0(|\vec{q}_\perp-\vec{k}_\perp| \tau) Y_1(k_\perp \tau)
	= - \frac{1}{\pi} \frac{1}{q_\perp k_\perp}\;, \\
	&&\int d \tau \tau J_1(q_\perp \tau) J_1(|\vec{q}_\perp-\vec{k}_\perp| \tau) Y_0(k_\perp \tau)
	=  \frac{1}{\pi} \frac{1}{q_\perp |\vp{k}-\vp{q}|}\;, \\
	&&\int d\tau \tau
	J_1(q_\perp \tau)  J_1(k_\perp \tau) J_0(|\vp{k}-\vp{q}|\tau)
	=
	\frac{1}{\pi q k} \frac{\vp{q} \cdot \vp{k} }{|\vp{q} \times \vp{k}|}\;,
	\\
	&&\int d\tau \tau
	J_1(q_\perp \tau)  J_1(|\vp{k}-\vp{q}| \tau) J_0(k_\perp \tau)
	\notag \\ &&\quad \quad  =
	\frac{1}{\pi q |\vp{k}-\vp{q}| } \frac{\vp{q} \cdot (\vp{q}-\vp{k}) }{|\vp{q} \times \vp{k}|}
	\;.
\end{eqnarray}
For completeness we also list the limits used in the main text
\begin{align}
	&\lim_{\tau \to 0 } \frac{J_1(k_\perp \tau)}{k_\perp \tau} = \frac12\;,\\
	&\lim_{\tau\to 0 } \tau \partial_\tau H_0^{(1,2)} (k_\perp \tau) =
	\pm i \frac{2}{\pi}\;,  \\
	&\lim_{\tau\to 0 } \tau^3 \partial_\tau \tau^{-1} H_1^{(1,2)} (k_\perp \tau) =
	\pm i \frac{4}{\pi k_\perp}\;
\end{align}
and the identity connecting the transverse projector and antisymmetric symbols
\begin{equation}
	\label{Eq:t_to_e}
	a_i b_j \mathfrak{t}_{ij}(\vp{k}) =
	\frac{\epsilon_{ij} k_j a_i}{k_\perp}
	\frac{\epsilon_{nm} k_n b_m}{k_\perp} \;.
\end{equation}

\section{Appendix C: Average with respect to  projectile configurations in MV model} \label{appC}
Lets consider the following combination averaged with respect to the
projectile field in the MV model
\begin{align}
\Omega_{ij,lm}^{a,b} (\vp{p},\vp{q}) & \equiv
\langle
\Omega_{ij}^a (\vp{p})
\Omega_{lm}^b (\vp{q})
\rangle _{\rho_1}
= \notag \\
& = g^2
\int \frac{d^2 u} {(2\pi)^2}
\int \frac{d^2 v} {(2\pi)^2}
\frac{ u_i (p-u)_j v_l (q-v)_m }{u_\perp^2 v_\perp^2}
\langle
\rho^\alpha_1(\vp{u}) \rho^\beta_1(\vp{v})
\rangle_{\rho_1} \notag \\ &
\quad \times W_{a \alpha}(\vp{p}-\vp{u})  W_{b \beta}(\vp{q}-\vp{v}) = \notag \\
& = g^2
\int \frac{d^2 u} {(2\pi)^2}
\mu^2(\vp{u})
\frac{ u_i (u+p)_j u_l (u-q)_m }{u_\perp^4}
\notag \\
& \quad \times
\left[
W(\vp{u}+\vp{p})
W^\dagger(\vp{u}-\vp{q})
\right]_{ab},
\label{Eq:Omega_double}
\end{align}
where we used the MV correlator
\begin{equation}
\langle
\rho^\alpha_1(\vp{u}) \rho^\beta_1(\vp{v})
\rangle_{\rho_1}
= (2\pi)^2 \mu^2(\vp{u})  \delta(\vp{u}+\vp{v}).
\end{equation}
We also use the notation from Sec.~\ref{Sec:DIP}
\begin{align}
f(\vp{k}) &=
		\frac{2g}{(2\pi)^3 k_\perp^2}
		\int \frac{d^2 q}{(2\pi)^2}
		\frac{\vp{k}\times \vp{q}}{|\vp{k}\times \vp{q}|}
		\frac{1}{q_\perp^2 |\vp{k}-\vp{q}|^2 }
		\notag \\
		&\times
		f^{abc}
		\Omega^a_{ij} (\vp{q})
		\Omega^b_{mn} (\vp{k}-\vp{q})
		\Omega^{c\star}_{rp} (\vp{k})
		\notag
		\times  \\ &
		\left[
			\left(
			k_\perp^2 \epsilon^{ij} \epsilon^{mn}
		-\vp{q} \cdot (\vp{k} - \vp{q} )
		(\epsilon^{ij} \epsilon^{mn}+\delta^{ij} \delta^{mn})
		\right) \epsilon^{rp}
		\right.
		\notag \\ &
		\left.
		+
		2 \vp{k} \cdot (\vp{k}-\vp{q}) {\epsilon^{ij} \delta^{mn}} \delta^{rp}
		\right].
\label{Eq:f}
\end{align}
Using Eq.~\eqref{Eq:Omega_double} we can rewrite
the projectile averaged combination $\langle f(\vp{p}) f(\vp{k}) \rangle_{\rho_1}$ as
follows
\begin{align}
&\langle f(\vp{p}) f(\vp{k}) \rangle_{\rho_1} =
\left(\frac{2g}{(2\pi)^3}\right)^2
\frac{1}{p_\perp^2  k_\perp^2}
\int \frac{d^2 q}{(2\pi)^2}
\int \frac{d^2 q'}{(2\pi)^2}
	\frac{\vp{p}\times \vp{q}}{|\vp{p}\times \vp{q}|}
	\frac{\vp{k}\times \pvp{q}}{|\vp{k}\times \pvp{q}|}
	\times
	\notag \\ &
	\frac{f^{abc}}{q_\perp^2 |\vp{p}-\vp{q}|^2 }
	\frac{f^{a'b'c'}}{ {q'_\perp}^2 |\vp{k}-{\pvp{q}}|^2 } \notag
\times
\\
&
\left[
			\left(
			p_\perp^2 \epsilon^{ij} \epsilon^{mn}
		-\vp{q} \cdot (\vp{p} - \vp{q} )
		(\epsilon^{ij} \epsilon^{mn}+\delta^{ij} \delta^{mn})
		\right) \epsilon^{rp}
		\right. \notag \\
		&\left.  \quad
		+ 2 \vp{p} \cdot (\vp{p}-\vp{q}) { \epsilon^{ij} \delta^{mn}} \delta^{rp}
		\right]
		\notag \times \\
&\left[
			\left(
			k_\perp^2 \epsilon^{i'j'} \epsilon^{m'n'}
		-\pvp{q} \cdot (\vp{k} - \pvp{q} )
		(\epsilon^{i'j'} \epsilon^{m'n'}+\delta^{i'j'} \delta^{m'n'})
		\right) \epsilon^{r'p'}
			\right. \notag \\
		&\left.  \quad
		+ 2 \vp{k} \cdot (\vp{k}-\pvp{q}) { \epsilon^{i'j'} \delta^{m'n'} } \delta^{r'p'}
		\right] \times \notag \\
			&
		\Big[
			\Omega^{a,b}_{ij,mn}(\vp{q},\vp{k}-\vp{q}) \left(
		\Omega^{c,a'}_{rp,i'j'}(-\vp{p},\pvp{q})
		\Omega^{b',c'}_{m'n',r'p'}(\vp{k}-\pvp{q},-\vp{k}) +\right.   \notag \\ & \left.
		\Omega^{c,b'}_{rp,m'n'}(-\vp{p},\vp{k}-\pvp{q})\Omega^{a',c'}_{i'j',r'p'}(\pvp{q},-\vp{k}) +
\right.  \notag \\ & \left.
		\Omega^{c,c'}_{rp,r'p'}(-\vp{p},-\vp{k})\Omega^{a',b'}_{i'j',m'n'}(\pvp{q},\vp{k}-\pvp{q})
		\right)   \notag \\
		&
		\Omega^{a,c}_{ij,rp}(\vp{q},-\vp{p})
		\left(
		\Omega^{b,a'}_{mn,i'j'}(\vp{p}-\vp{p},\pvp{q}) \Omega^{b',c'}_{m'n',r'p'}(\vp{k}-\pvp{q},-\vp{k}) + \right. \notag \\ & \left.
		\Omega^{b,b'}_{mn,m'n'}(\vp{p}-\vp{q},\vp{k}-\pvp{q})\Omega^{a',c'}_{i'j',r'p'}(\pvp{q},-\vp{k}) +
\right.  \notag \\ &  \left.
		\Omega^{b,c'}_{mn,r'p'}(\vp{p}-\vp{q},-\vp{k})\Omega^{a',b'}_{i'j',m'n'}(\vp{q}',\vp{k}-\pvp{q})
		\right)  +
		\notag \\
	&
		\Omega^{a,a'}_{ij,i'j'}(\vp{q},\pvp{q})
		\left(
		\Omega^{b,c}_{mn,rp}(\vp{p}-\vp{p},-\vp{p}) \Omega^{b',c'}_{m'n',r'p'}(\vp{k}-\pvp{q},-\vp{k}) + \right. \notag \\ & \left.
		\Omega^{b,b'}_{mn,m'n'}(\vp{p}-\vp{q},\vp{k}-\pvp{q})\Omega^{c,c'}_{rp,r'p'}(-\vp{p},-\vp{k}) +
\right.  \notag \\ &  \left.
		\Omega^{b,c'}_{mn,r'p'}(\vp{p}-\vp{q},-\vp{k})\Omega^{c,b'}_{rp,m'n'}(-\vp{p}',\vp{k}-\pvp{q})
		\right)  +
		\notag \\
		&
		\Omega^{a,b'}_{ij,m'n'}(\vp{q},\vp{k}-\pvp{q})
		\left(
		\Omega^{b,c}_{mn,rp}(\vp{p}-\vp{p},-\vp{p}) \Omega^{a',c'}_{i'j',r'p'}(\pvp{q},-\vp{k}) + \right. \notag \\ & \left.
		\Omega^{b,a'}_{mn,i'j'}(\vp{p}-\vp{q},\pvp{q}) \Omega^{c,c'}_{rp,r'p'}(-\vp{p},-\vp{k}) +
\right.  \notag \\ &  \left.
		\Omega^{b,c'}_{mn,r'p'}(\vp{p}-\vp{q},-\vp{k})
		\Omega^{c,a'}_{rp,i'j'}(-\pvp{p},\pvp{q})
		\right)  +
		\notag \\
		&
		\Omega^{a,c'}_{ij,r'p'}(\vp{q},-\vp{k})
		\left(
		\Omega^{b,c}_{mn,rp}(\vp{p}-\vp{p},-\vp{p}) \Omega^{a',b'}_{i'j',m'n'}(\pvp{q},\vp{k}-\pvp{q}) + \right. \notag \\ & \left.
		\Omega^{b,a'}_{mn,i'j'}(\vp{p}-\vp{q},\pvp{q}) \Omega^{c,b'}_{rp,m'n'}(-\vp{p},\vp{k}-\pvp{q}) +
\right.  \notag \\ &  \left.
		\Omega^{b,b'}_{mn,m'n'}(\vp{p}-\vp{q},\vp{k}-\pvp{q})
		\Omega^{c,a'}_{rp,i'j'}(-\pvp{p},\pvp{q})
	\right)\Big]\,.
		\label{Eq:f_av}
	\end{align}


\begin{thebibliography} {000}







\bibitem{Dumitru:2008wn} 
  A.~Dumitru, F.~Gelis, L.~McLerran and R.~Venugopalan,
  Nucl.\ Phys.\ A {\bf 810}, 91 (2008)
  doi:10.1016/j.nuclphysa.2008.06.012
  [arXiv:0804.3858 [hep-ph]].


\bibitem{Dumitru:2010iy} 
  A.~Dumitru, K.~Dusling, F.~Gelis, J.~Jalilian-Marian, T.~Lappi and R.~Venugopalan,
  Phys.\ Lett.\ B {\bf 697}, 21 (2011)
  doi:10.1016/j.physletb.2011.01.024
  [arXiv:1009.5295 [hep-ph]].


\bibitem{Dusling:2012iga} 
  K.~Dusling and R.~Venugopalan,
  Phys.\ Rev.\ Lett.\  {\bf 108}, 262001 (2012)
  doi:10.1103/PhysRevLett.108.262001
  [arXiv:1201.2658 [hep-ph]].


\bibitem{Dusling:2012wy} 
  K.~Dusling and R.~Venugopalan,
  Phys.\ Rev.\ D {\bf 87}, no. 5, 054014 (2013)
  doi:10.1103/PhysRevD.87.054014
  [arXiv:1211.3701 [hep-ph]].


\bibitem{Alver:2006wh} 
  B.~Alver {\it et al.} [PHOBOS Collaboration],
  Phys.\ Rev.\ Lett.\  {\bf 98}, 242302 (2007)
  doi:10.1103/PhysRevLett.98.242302
  [nucl-ex/0610037].


\bibitem{Alver:2008zza} 
  B.~Alver {\it et al.} [PHOBOS Collaboration],
  Phys.\ Rev.\ C {\bf 77}, 014906 (2008)
  doi:10.1103/PhysRevC.77.014906
  [arXiv:0711.3724 [nucl-ex]].


\bibitem{Bzdak:2013rya} 
  A.~Bzdak, P.~Bozek and L.~McLerran,
  Nucl.\ Phys.\ A {\bf 927}, 15 (2014)
  doi:10.1016/j.nuclphysa.2014.03.007
  [arXiv:1311.7325 [hep-ph]].


\bibitem{Yan:2014afa} 
  L.~Yan, J.~Y.~Ollitrault and A.~M.~Poskanzer,
  Phys.\ Rev.\ C {\bf 90}, no. 2, 024903 (2014)
  doi:10.1103/PhysRevC.90.024903
  [arXiv:1405.6595 [nucl-th]].


\bibitem{Bzdak:2013raa} 
  A.~Bzdak and V.~Skokov,
  Nucl.\ Phys.\ A {\bf 943}, 1 (2015)
  doi:10.1016/j.nuclphysa.2015.08.001
  [arXiv:1312.7349 [hep-ph]].


\bibitem{Dumitru:2014yza} 
  A.~Dumitru, L.~McLerran and V.~Skokov,
  Phys.\ Lett.\ B {\bf 743}, 134 (2015)
  doi:10.1016/j.physletb.2015.02.046
  [arXiv:1410.4844 [hep-ph]].


\bibitem{McLerran:2015sva} 
  L.~McLerran and V.~Skokov,
  Nucl.\ Phys.\ A {\bf 947}, 142 (2016)
  doi:10.1016/j.nuclphysa.2015.12.005
  [arXiv:1510.08072 [hep-ph]].


\bibitem{McLerran:2016ivs} 
  L.~McLerran and V.~V.~Skokov,
  Nucl.\ Phys.\ A {\bf 957}, 230 (2017)
  doi:10.1016/j.nuclphysa.2016.09.004
  [arXiv:1604.05286 [hep-ph]].


\bibitem{Skokov:2014tka} 
  V.~Skokov,
  Phys.\ Rev.\ D {\bf 91}, no. 5, 054014 (2015)
  doi:10.1103/PhysRevD.91.054014
  [arXiv:1412.5191 [hep-ph]].


\bibitem{Dumitru:2015cfa} 
  A.~Dumitru, A.~V.~Giannini and V.~Skokov,
  arXiv:1503.03897 [hep-ph].




\bibitem{Gale:2012rq} 
  C.~Gale, S.~Jeon, B.~Schenke, P.~Tribedy and R.~Venugopalan,
  Phys.\ Rev.\ Lett.\  {\bf 110}, no. 1, 012302 (2013)
  doi:10.1103/PhysRevLett.110.012302
  [arXiv:1209.6330 [nucl-th]].


\bibitem{Lappi:2015vta} 
  T.~Lappi, B.~Schenke, S.~Schlichting and R.~Venugopalan,
  JHEP {\bf 1601}, 061 (2016)
  doi:10.1007/JHEP01(2016)061
  [arXiv:1509.03499 [hep-ph]].


\bibitem{Schenke:2015aqa} 
  B.~Schenke, S.~Schlichting and R.~Venugopalan,
  Phys.\ Lett.\ B {\bf 747}, 76 (2015)
  doi:10.1016/j.physletb.2015.05.051
  [arXiv:1502.01331 [hep-ph]].


\bibitem{Kovner:1995ja} 
  A.~Kovner, L.~D.~McLerran and H.~Weigert,
  Phys.\ Rev.\ D {\bf 52}, 6231 (1995)
  doi:10.1103/PhysRevD.52.6231
  [hep-ph/9502289].


\bibitem{Kovner:1995ts} 
  A.~Kovner, L.~D.~McLerran and H.~Weigert,
  Phys.\ Rev.\ D {\bf 52}, 3809 (1995)
  doi:10.1103/PhysRevD.52.3809
  [hep-ph/9505320].


\bibitem{Kovner:2010xk} 
  A.~Kovner and M.~Lublinsky,
  Phys.\ Rev.\ D {\bf 83}, 034017 (2011)
  doi:10.1103/PhysRevD.83.034017
  [arXiv:1012.3398 [hep-ph]].


\bibitem{Kovner:2012jm} 
  A.~Kovner and M.~Lublinsky,
  Int.\ J.\ Mod.\ Phys.\ E {\bf 22}, 1330001 (2013)
  doi:10.1142/S0218301313300014
  [arXiv:1211.1928 [hep-ph]].


\bibitem{Lappi:2009xa} 
  T.~Lappi, S.~Srednyak and R.~Venugopalan,
  JHEP {\bf 1001}, 066 (2010)
  doi:10.1007/JHEP01(2010)066
  [arXiv:0911.2068 [hep-ph]].


\bibitem{Gelis:2010nm} 
  F.~Gelis, E.~Iancu, J.~Jalilian-Marian and R.~Venugopalan,
  Ann.\ Rev.\ Nucl.\ Part.\ Sci.\  {\bf 60}, 463 (2010)
  doi:10.1146/annurev.nucl.010909.083629
  [arXiv:1002.0333 [hep-ph]].


\bibitem{Dumitru:2001ux} 
  A.~Dumitru and L.~D.~McLerran,
  Nucl.\ Phys.\ A {\bf 700}, 492 (2002)
  doi:10.1016/S0375-9474(01)01301-X
  [hep-ph/0105268].


\bibitem{Gelis:2008ad} 
  F.~Gelis, T.~Lappi and R.~Venugopalan,
  Phys.\ Rev.\ D {\bf 78}, 054020 (2008)
  doi:10.1103/PhysRevD.78.054020
  [arXiv:0807.1306 [hep-ph]].


\bibitem{Chirilli:2015tea} 
  G.~A.~Chirilli, Y.~V.~Kovchegov and D.~E.~Wertepny,
  JHEP {\bf 1503}, 015 (2015)
  doi:10.1007/JHEP03(2015)015
  [arXiv:1501.03106 [hep-ph]].


\bibitem{Kovchegov:2018jun} 
  Y.~V.~Kovchegov and V.~V.~Skokov,
  Phys.\ Rev.\ D {\bf 97}, no. 9, 094021 (2018)
  doi:10.1103/PhysRevD.97.094021
  [arXiv:1802.08166 [hep-ph]].


\bibitem{Mace:2018vwq} 
  M.~Mace, V.~V.~Skokov, P.~Tribedy and R.~Venugopalan,
  arXiv:1805.09342 [hep-ph].


\bibitem{Kovner:2016jfp} 
  A.~Kovner, M.~Lublinsky and V.~Skokov,
  Phys.\ Rev.\ D {\bf 96}, no. 1, 016010 (2017)
  doi:10.1103/PhysRevD.96.016010
  [arXiv:1612.07790 [hep-ph]].


\bibitem{Gale:2012rq} 
  C.~Gale, S.~Jeon, B.~Schenke, P.~Tribedy and R.~Venugopalan,
  Phys.\ Rev.\ Lett.\  {\bf 110}, no. 1, 012302 (2013)
  doi:10.1103/PhysRevLett.110.012302
  [arXiv:1209.6330 [nucl-th]].


\bibitem{Lappi:2015vta} 
  T.~Lappi, B.~Schenke, S.~Schlichting and R.~Venugopalan,
  JHEP {\bf 1601}, 061 (2016)
  doi:10.1007/JHEP01(2016)061
  [arXiv:1509.03499 [hep-ph]].


\bibitem{Schenke:2015aqa} 
  B.~Schenke, S.~Schlichting and R.~Venugopalan,
  Phys.\ Lett.\ B {\bf 747}, 76 (2015)
  doi:10.1016/j.physletb.2015.05.051
  [arXiv:1502.01331 [hep-ph]].


\bibitem{Kovner:1995ja} 
  A.~Kovner, L.~D.~McLerran and H.~Weigert,
  Phys.\ Rev.\ D {\bf 52}, 6231 (1995)
  doi:10.1103/PhysRevD.52.6231
  [hep-ph/9502289].


\bibitem{Kovner:1995ts} 
  A.~Kovner, L.~D.~McLerran and H.~Weigert,
  Phys.\ Rev.\ D {\bf 52}, 3809 (1995)
  doi:10.1103/PhysRevD.52.3809
  [hep-ph/9505320].


\bibitem{Kovner:2010xk} 
  A.~Kovner and M.~Lublinsky,
  Phys.\ Rev.\ D {\bf 83}, 034017 (2011)
  doi:10.1103/PhysRevD.83.034017
  [arXiv:1012.3398 [hep-ph]].


\bibitem{Kovner:2012jm} 
  A.~Kovner and M.~Lublinsky,
  Int.\ J.\ Mod.\ Phys.\ E {\bf 22}, 1330001 (2013)
  doi:10.1142/S0218301313300014
  [arXiv:1211.1928 [hep-ph]].


\bibitem{Lappi:2009xa} 
  T.~Lappi, S.~Srednyak and R.~Venugopalan,
  JHEP {\bf 1001}, 066 (2010)
  doi:10.1007/JHEP01(2010)066
  [arXiv:0911.2068 [hep-ph]].


\bibitem{Gelis:2010nm} 
  F.~Gelis, E.~Iancu, J.~Jalilian-Marian and R.~Venugopalan,
  Ann.\ Rev.\ Nucl.\ Part.\ Sci.\  {\bf 60}, 463 (2010)
  doi:10.1146/annurev.nucl.010909.083629
  [arXiv:1002.0333 [hep-ph]].


\bibitem{Dumitru:2001ux} 
  A.~Dumitru and L.~D.~McLerran,
  Nucl.\ Phys.\ A {\bf 700}, 492 (2002)
  doi:10.1016/S0375-9474(01)01301-X
  [hep-ph/0105268].


\bibitem{Gelis:2008ad} 
  F.~Gelis, T.~Lappi and R.~Venugopalan,
  Phys.\ Rev.\ D {\bf 78}, 054020 (2008)
  doi:10.1103/PhysRevD.78.054020
  [arXiv:0807.1306 [hep-ph]].


\bibitem{Chirilli:2015tea} 
  G.~A.~Chirilli, Y.~V.~Kovchegov and D.~E.~Wertepny,
  JHEP {\bf 1503}, 015 (2015)
  doi:10.1007/JHEP03(2015)015
  [arXiv:1501.03106 [hep-ph]].


\bibitem{Kovchegov:2018jun} 
  Y.~V.~Kovchegov and V.~V.~Skokov,
  Phys.\ Rev.\ D {\bf 97}, no. 9, 094021 (2018)
  doi:10.1103/PhysRevD.97.094021
  [arXiv:1802.08166 [hep-ph]].


\bibitem{Mace:2018vwq} 
  M.~Mace, V.~V.~Skokov, P.~Tribedy and R.~Venugopalan,
  arXiv:1805.09342 [hep-ph].


\bibitem{prudnikov_integrals_1998}
  A.~P.~Prudnikov and O.~I. Marichev, {\emph {Integrals and
  Series: Special functions}}\ ({CRC Press},\
  {1998}).

\end{thebibliography}
\end{document}